\documentclass{llncs}
\usepackage[T1]{fontenc}
\usepackage{graphicx}
\usepackage{algorithmic}
\usepackage{textcomp}
\usepackage{algorithm2e}
\usepackage{xcolor}
\usepackage{subcaption}
\begin{document}
\title{A DirectX-Based DICOM Viewer for Multi-User Surgical Planning in Augmented Reality\thanks{This research was funded in part by a grant from The Leona M. and Harry B. Helmsley  Charitable  Trust. The authors would also like to thank the volunteers who provided their medical data for us to work with under UCSD IRB protocol $\#182146$.}}

% \titlerunning{Abbreviated paper title}
% If the paper title is too long for the running head, you can set
% an abbreviated paper title here

\author{Menghe Zhang\and
Weichen Liu\and
Nadir Weibel \and
J\"urgen P. Schulze}

% \authorrunning{F. Author et al.}
% First names are abbreviated in the running head.
% If there are more than two authors, 'et al.' is used.

\institute{University of California, San Diego\\ 
\email{\{mez071, wel008, weibel, jschulze\}@ucsd.edu}\\}

\maketitle              % typeset the header of the contribution
\begin{abstract}
Preoperative medical imaging is an essential part of surgical planning. The data from medical imaging devices, such as CT and MRI scanners, consist of stacks of 2D images in DICOM format. Conversely, advances in 3D data visualization provide further information by assembling cross-sections into 3D volumetric datasets. As Microsoft unveiled the HoloLens 2 (HL2), which is considered one of the best Mixed Reality (XR) headsets in the market, it promised to enhance visualization in 3D by providing an immersive experience to users. This paper introduces a prototype holographic XR DICOM Viewer for the 3D visualization of DICOM image sets on HL2 for surgical planning. We first developed a standalone graphical C++ engine using the native DirectX11 API and HLSL shaders. Based on that, the prototype further applies the OpenXR API for potential deployment on a wide range of devices from vendors across the XR spectrum. With native access to the device, our prototype unravels the limitation of hardware capabilities on HL2 for 3D volume rendering and interaction. Moreover, smartphones can act as input devices to provide another user interaction method by connecting to our server. In this paper, we present a holographic DICOM viewer for the HoloLens 2 and contribute (i) a prototype that renders the DICOM image stacks in real-time on HL2, (ii) three types of user interactions in XR, and (iii) a preliminary qualitative evaluation of our prototype.

\end{abstract}

\section{Introduction}
Surgical planning with medical imaging in a DICOM format is a demanding process. Traditionally, viewing and planning on 2D screens make it challenging for clinical practitioners to extract 3D information such as anatomical structures, the regions of interest, and spatial relationships among the image set. 

3D volumetric visualization in Mobile Augmented Reality (MAR) solves this problem by integrating the volumetric object with the physical environment for mobile devices. Our previous work on Android\cite{zhang2021server} developed an application to load, view, and interact with 3D medical volumes in traditional 2D screen and AR scenarios. 

\begin{figure}
  \centering
  \includegraphics[width=0.9\columnwidth]{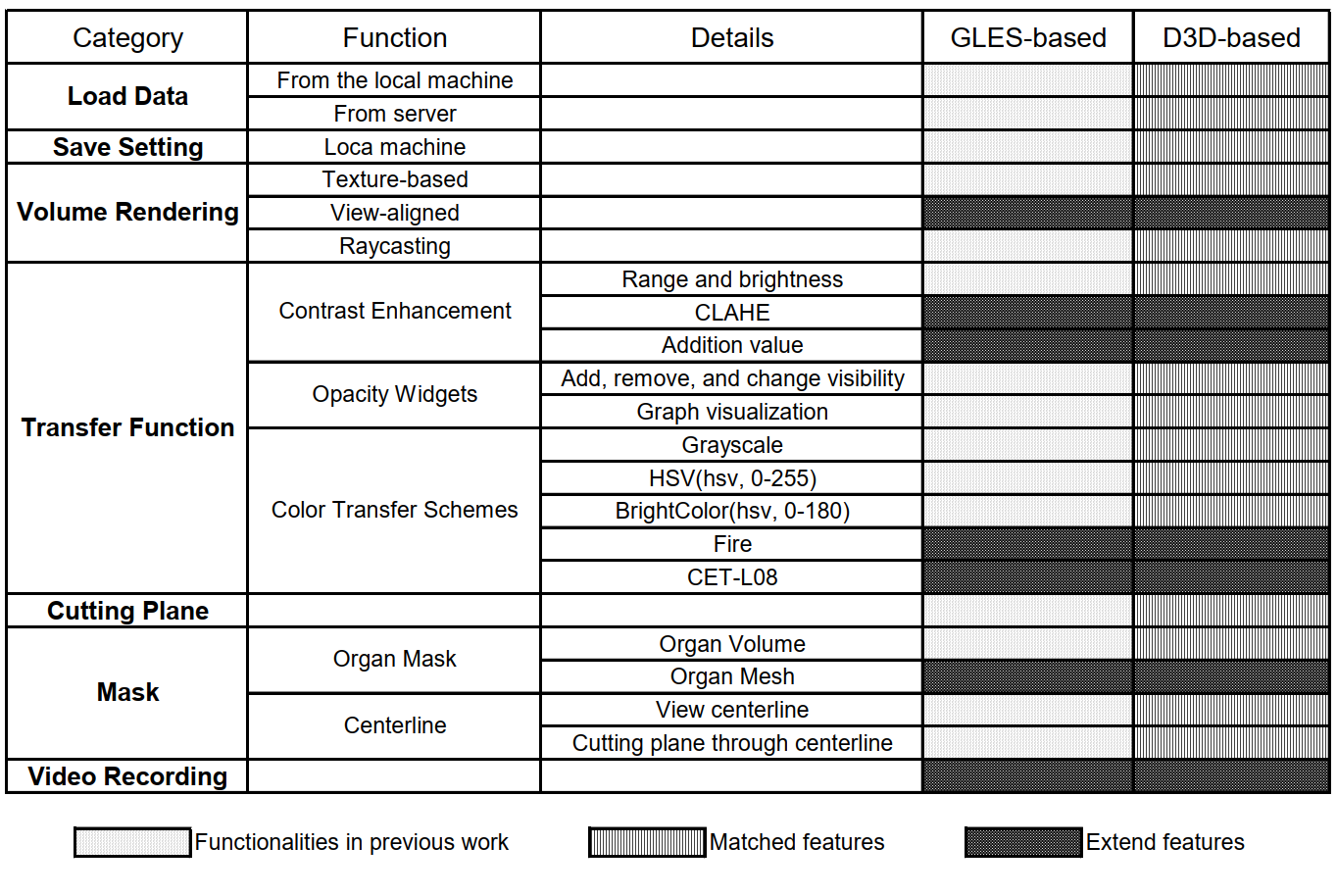} 
  \caption{List of the functionalities on GLES-based and D3D-based applications.}~\label{fig:feature-list}
  \vspace{-1.5em}
\end{figure}

To further allow hands-free operations, we adapt it to HoloLens 2 with our graphical C++ MR rendering engine using the native DirectX11 API and HLSL shaders to boost the device's performance. We also show three user interaction methodologies for different tasks: 1) we design our own interaction set (3D interactions) and interfaces (virtual 2D plane touch) for unique user interactions using OpenXR APIs; 2) with matched functionalities to our Android DICOM Viewer, we make Android phones as input devices to MR applications. Fig.~\ref{fig:feature-list} shows the categorized feature list. We will detail the implementation of the extended features in this section.

%-------------------------------------------------------------------------
\section{Related Work}
\subsection{Holographic DICOM Viewer Prototypes}
Mojica et al.~\cite{mojica2017holographic} proposed an AR DICOM Viewer prototype for the first generation of the HoloLens. To accelerate volume rendering, they only render the plane under a specific setting to avoid rendering the whole data volume. Remote rendering is another solution. Frohlich et al.~\cite{frohlich2018holographic} proposed it as part of a dedicated workstation with a wireless connection that enables real-time computation on MR headsets. Their latest work~\cite{morales2021holographic} is deployed in a two-CPU fashion: parts run on the HoloLens and parts run on an external PC connected through Wi-Fi to enable real-time experience.

HoloDICOM~\cite{holodicom} is a standard DICOM image stack visualization tool targets at the HoloLens 2. It contains tools for volume manipulation, a 2D slicer, and an intuitive floating 2D user interface. However, only one instructor can interact with the application, while others could only observe the interfaces. 

%-------------------------------------------------------------------------
\subsection{Interaction with 3D Objects}
3D selection in mid-air is commonly used in XR applications. However, interacting with far away objects and dense object selection still create problems in terms of user fatigue~\cite{jang2017modeling}. We have been exploring to apply 2D touch to achieve 3D selection in AR~\cite{zhang2019arcalvr}\cite{zhang2020calar}

The idea of tuning smartphones info input devices for HMDs attracted the designers' interest in various interactions. Karan et al.~\cite{karan2021smartphone} enhanced the AR game experience by creating new interaction methods through a smartphone used as a game controller. Their remote input SDK makes connections between Unity applications developed for XR devices. Buschel et al.~\cite{buschel2019investigating} contributes an exploration of smartphone-based interaction techniques for 3D pan and zoom, which are two essential tasks in HMD applications. They designed five 2D touch screen gestures mapped to pan $\&$ zoom operations in 3D. MARVIS~\cite{langner2021marvis} combines smartphones and HMD AR for data analysis. 

Until now, few medical imaging-based surgical planning solutions on the HoloLens take full advantage of the flexibilities of the native engine.
\begin{figure}[ht!]
  \centering
  \includegraphics[width=0.8\columnwidth]{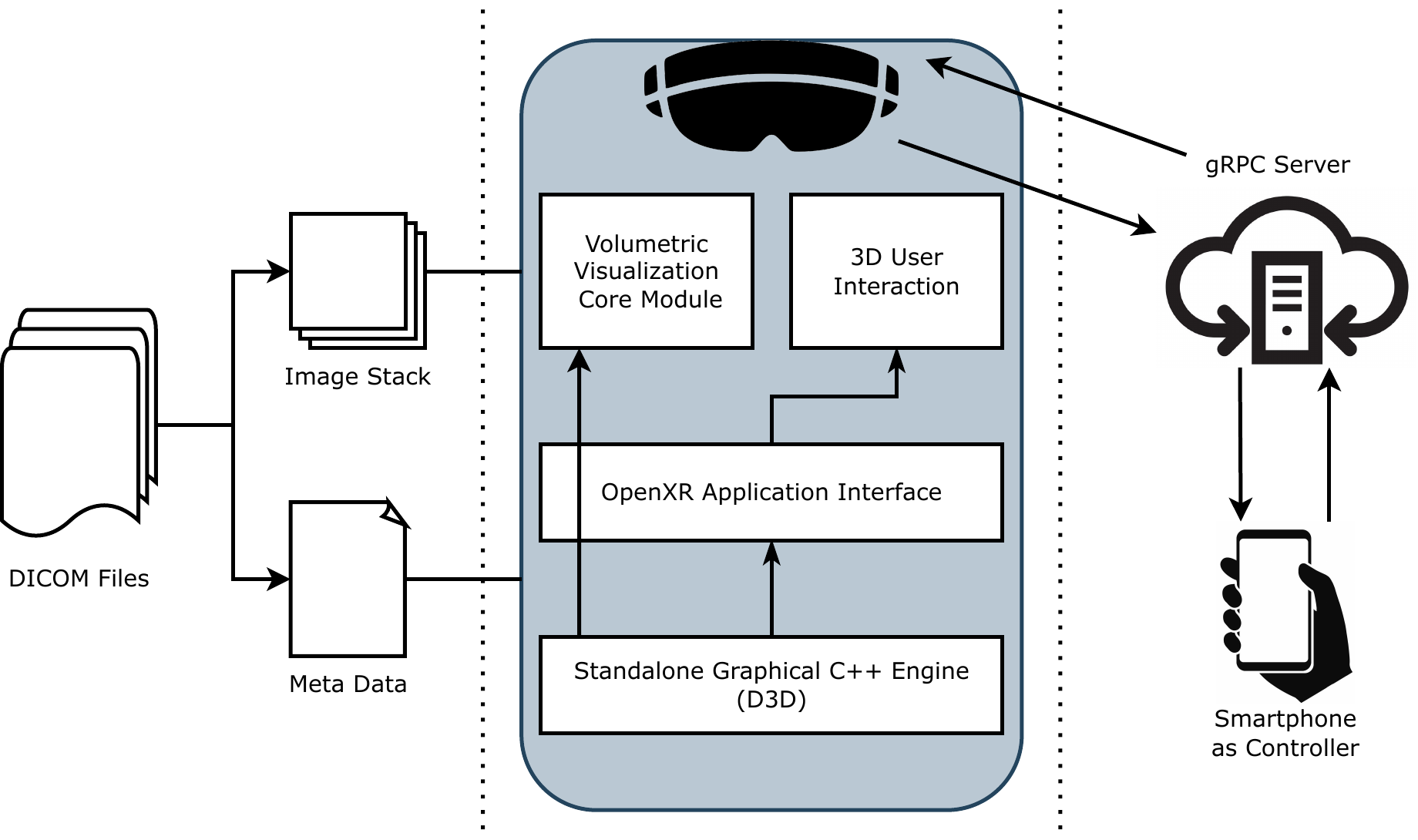}
  \caption{System structure: Offline DICOM data preprocessing and anonymization (left), Direct3D based DICOM viewer prototype structure (middle), and smartphone as input device (right)}~\label{fig:stru1}
\end{figure}

%------------------------------------------------------------------------
\section{System Design Overview}
\begin{figure}
  \centering
  \includegraphics[width=0.8\columnwidth]{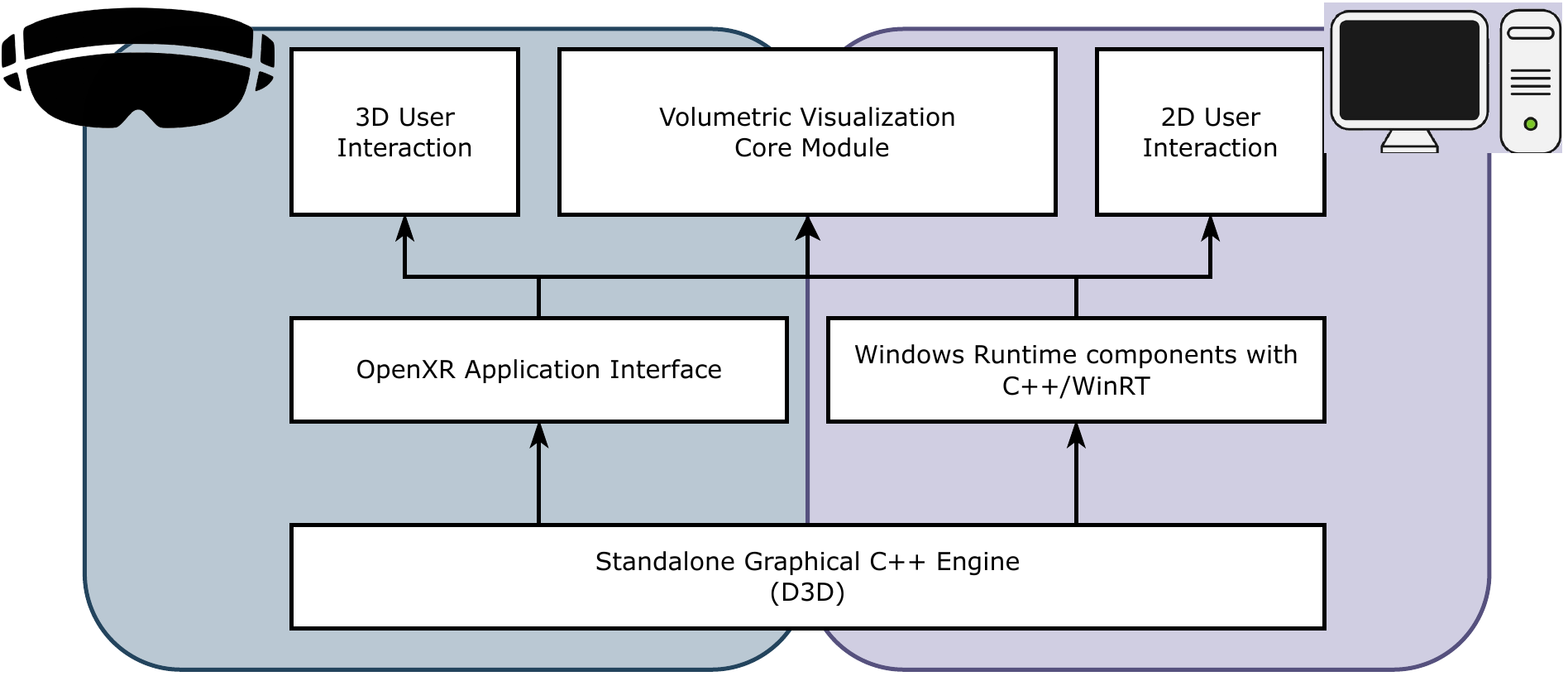}
  \caption{Direct3D based DICOM viewer on HoloLens 2 and desktop}~\label{fig:stru2}
\end{figure}

Fig.~\ref{fig:stru1} shows our system's components integrating an offline DICOM data preprocessor, a Direct3D-based DICOM Viewer, and the setup for smartphones as user input. Our previous work\cite{zhang2021server} introduces the DICOM data preprocessing tool on a desktop that categorizes information and anonymizes sensitive data. Then the graphical engine based on UWP with DirectX 11 API allows developers to manage hardware resources efficiently. The 3D renderer with OpenXR enables potential deployment on a wide range of MR devices.
On the project structure side, as HL2 moved to a 64-bit Windows 10 machine that runs UWP applications, our prototype branches into two applications: one for HL2, the other running on a Windows 10 desktop. Fig.~\ref{fig:stru2} illustrates the relationship between the two applications.

\section{Direct3D-based DICOM Viewer Implementation}

\subsection{Smartphones as User Input Devices}
One of the advantages of setting up the gRPC server in a local area network is that either HL2 users or other users can more flexibly input and control the AR application. As described in previous work~\cite{zhang2021server}, such a server is implemented with a gRPC framework that can run in any environment. We designed the functionalities on this D3D based DICOM Viewer to match those on a smartphone, following previous work~\cite{zhang2021server}, which allowed all operations on the smartphone to be translated and passed to the gRPC server seamlessly. 

%-------------------------------------------------------------------------

\subsection{Functionalities}

\subsubsection{Contrast Enhancement}
Previous work~\cite{zhang2021server} describes a uniform enhancement of contrast and brightness.

An extension of the contrast enhancement came from the fact that our DICOM data has a resolution of 12 bits per pixel. The strategy of compressing the original 12 bits for a typical 8-bit display reduces the image contrast. Normally, we would discard the last four digits to map the data to 8 bits, or use some linear mapping, but neither solves the problem of displaying these high dynamic range images on regular displays.

\begin{figure}
  \centering
  \includegraphics[width=0.9\columnwidth]{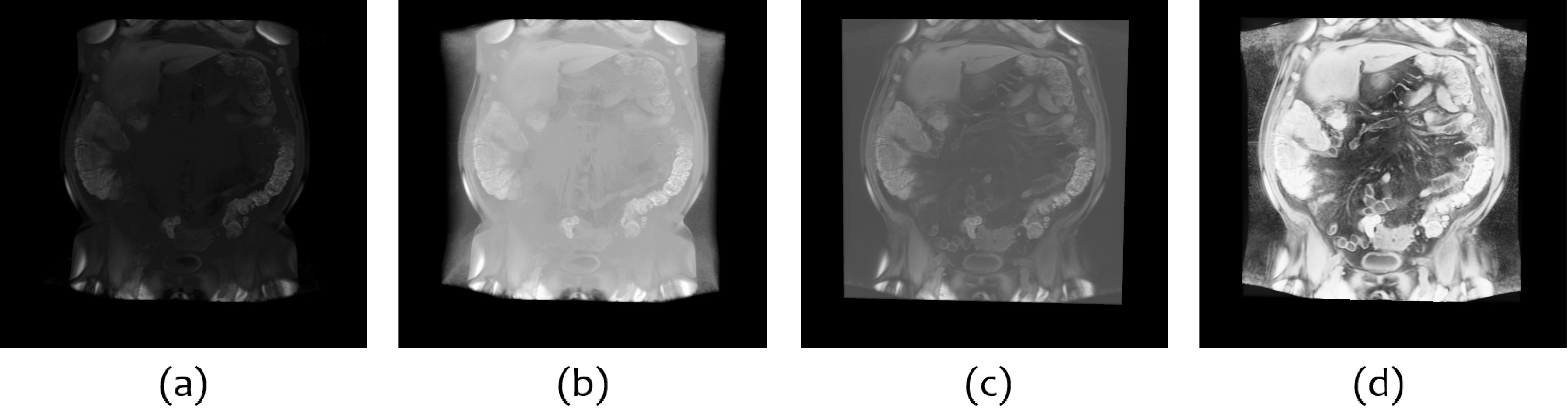} \caption{Comparison of different contrast enhancement methods: (a) raw data volume, (b) only increasing the brightness, (c) only increasing the base value, and (d) using CLAHE}~\label{fig:contrast-enhancement}
\end{figure}

Therefore, we chose to integrate Lucknavalai and Schulze's work~\cite{lucknavalai2020real} to perform real-time 3D Contrast Limited Adaptive Histogram Equalization (CLAHE) to enhance 3D medical image stacks. Fig.~\ref{fig:contrast-enhancement} compares the same volume data under the same setting only applies different contrast enhancement methods. Notice that for comparison, we only change one factor for each view. However, combined adjustments can be used to improve the quality further.

\subsubsection{Color Transfer Schemes}
\begin{figure}
  \centering
  \includegraphics[width=0.8\columnwidth]{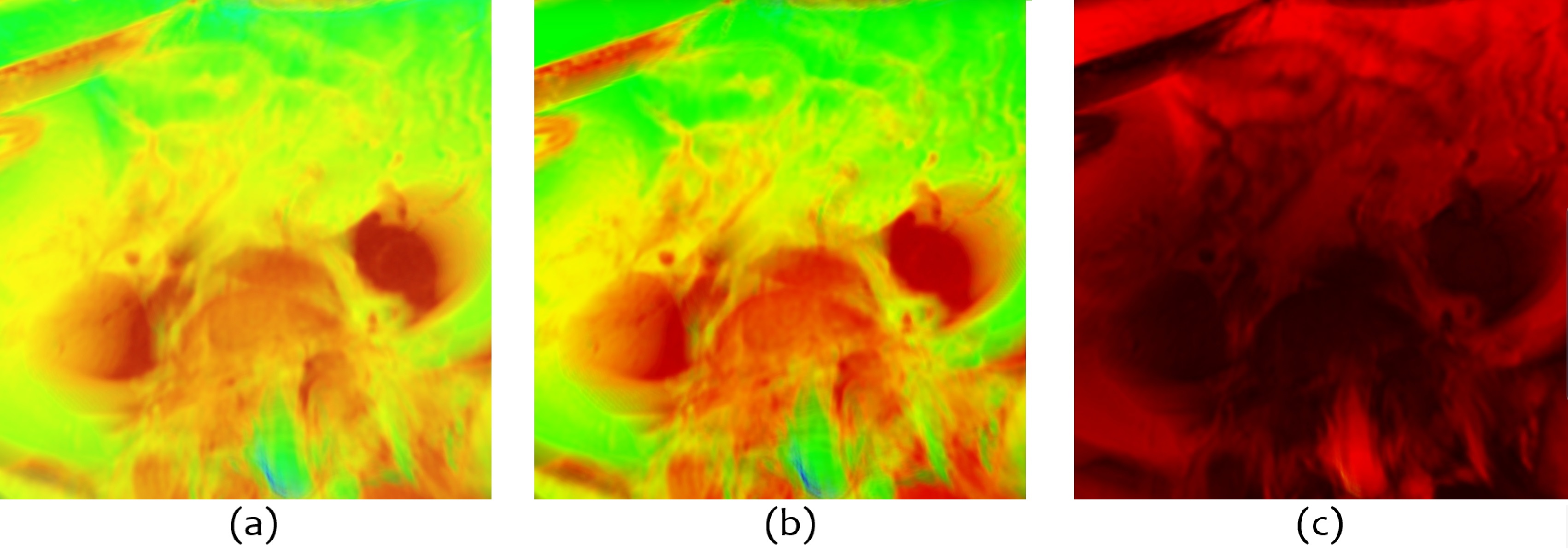} \caption{Tissue visualization under : (a) HSV color scheme, (b) fine-tuned HSV color scheme, and (c) the Fire color scheme}~\label{fig:color-scheme-detail}
\end{figure}

After presenting the color transfer schemes applied in previous work, we realized that the piecewise linear paths through RGB color space have two major issues. One of them is that the perceptual flat spots in the color map hide detailed features; the other is the perceptual discontinuities that induce false features. Fig.~\ref{fig:color-scheme-detail} (a) and (b) show a part of the volume rendered with the HSV color scheme. The clustering of points in the green and red regions produces perceptual flat spots in the color map, thus limiting the effects only by contrast enhancement and opacity adjustment. Also, yellow and red produce false anomalies seen at these spots. Therefore, we extend the color schemes with Fire and CET-L08 from Kovesi's work~\cite{kovesi2015good}, which uniform the magnitude of the incremental change in perceptual lightness. As shown in Fig.~\ref{fig:color-scheme-detail} (c), the two new color schemes perform better for detail perception compared to the previous ones. The figures also show how at fine spatial acuity, human eyes perform much better for luminance gratings than chromatic gratings.

\subsubsection{Video Recording}
We implemented a spectator view mode, which only renders the front RGB camera for live preview on the HoloLens 2. The initial motivation is that when recording for a demonstration video or streaming, the frame rate on the HoloLens 2 would decrease as it does the volumetric rendering for three different views per cycle. Due to the heavy computational intensity and limited pixel fill rate, the resulting low frame rate is not acceptable for interactive usage. Therefore, our solution only renders one view from the RGB camera and doubles the frame rate compared with regular use. 

HL2 provides a "Render from the PV camera" option on the Windows Device Portal. However, instead of handling all reprojection calculations and rendering the third view automatically, this option toggles a flag on the Hololens. It makes data coming from the PV camera available programmatically. We still need to enable corresponding OpenXR configs and manually handle the additional reprojected rendering for the PV camera. The data from the PV camera is self-contained enough to perform complete volumetric rendering individually without the need to do any computation from the regular display rendering. The rendered virtual hologram will then be blended with the RGB video captured by the PV camera to produce the final spectator view.

\subsection{Marker-based 3D Object Placement}
To achieve a marker-based 3D object placement and put multiple users under the same reference frame, we combine the marker tracking and the Hololens tracking. We integrate ArUco marker tracking~\cite{munoz2012aruco} with OpenCV~\cite{opencv2020} on the Hololens. Due to performance concerns, we implement the OpenCV thread parallel to the main rendering thread. The OpenCV builds the reference frame and tracks the printed ArUco marker based on the images from the Left-Front VLC camera every few rendering frames. Then we create/adjust a spatial anchor onto the detected ArUco marker and use it as the frame of the reference across all devices. 

OpenXR defines XrSpace to allow applications to track real-world objects. In our Hololens application, we create the following XrSpaces:

\begin{enumerate}
    \item Reference space: defines the original pose of the device.
    \item Hand pose space: defines in the hand tracking space locally for hand interaction
    \item Sensor space: a space provided by Research Mode \cite{researchmode2020} device sensor for coordinate alignments.
\end{enumerate}

Most of the spaces can be directly updated by querying OpenXR methods, except for sensor space, which is from Research Mode in the WinRT context. We utilize the sensor's rig node GUID with the Windows Perception APIs to acquire corresponding space representation in the OpenXR coordinate system.

\begin{figure}
\subfloat[]{%
\includegraphics[height=0.21\textheight]{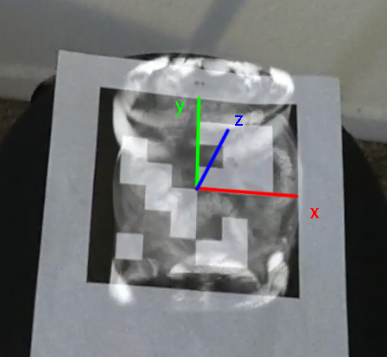}}
\hspace*{\fill}
\subfloat[]{%
\includegraphics[height=0.21\textheight]{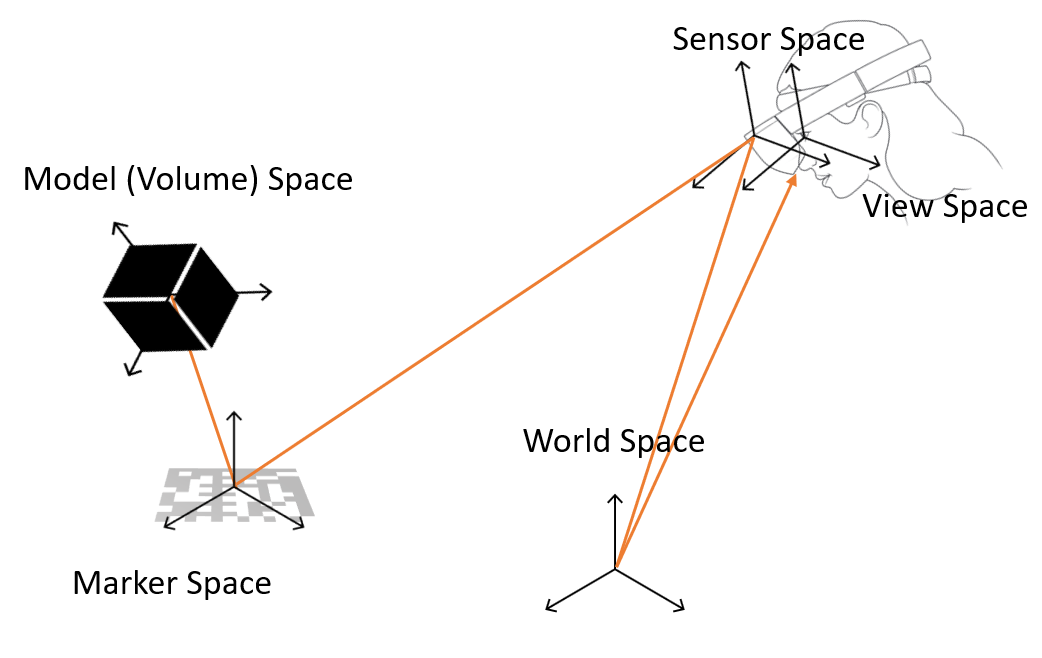}}
\caption{(a) Volume positioning with the ArUco Marker in Hololens; (b) The transformation map from volume space to view space with marker tracking.}
~\label{fig:marker-coordinate}
\end{figure}

Given the model matrix of the volume $ModelMatrix$, the world to view matrix $ViewMatrix$ and the projection matrix $ProjMatrix$ provided by the OpenXR Swapchain configuration, a volume is rendered by $ProjMatrix*ViewMatrix*ModelMatrix$ when it is not positioned by a ArUco marker. In this case, mid-air hand gestures control the volume's poses (scale, translation, and orientation). The hand movements in the HandSpace are first converted into world space, then multiplied by sensitivity value for finer control of the volume pose, and finally directly applied onto the $ModelMatrix$.

\begin{figure*}[ht!]
  \centering
  \includegraphics[width=0.9\columnwidth]{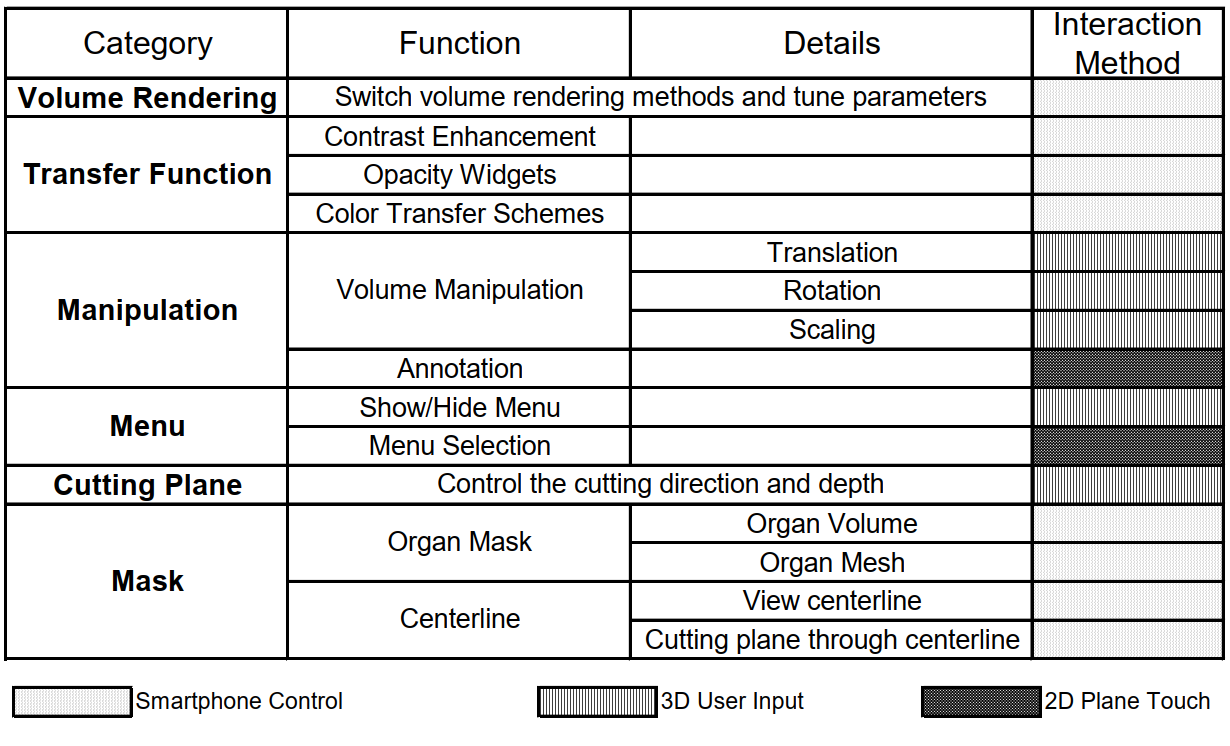} \caption{User interaction tasks with corresponding method.}~\label{fig:interaction-tasks}
\end{figure*}
We handle the case where we use the detected ArUco marker to position the volume differently. OpenCV detects the printed ArUco markers in the sensor's image and returns the marker's transform $Marker2Sensor$ with respect to the LF VLC camera. The sensor space representation in OpenXR provides us up-to-date sensor transform $Sensor2World$ in the world space. With this information, we can acquire the marker transform in the Hololens' world space by $Marker2World = Sensor2World * Marker2Sensor$. The volume is then rendered with the matrix transformation $ProjMatrix*ViewMatrix*Marker2World*ModelMatrix$ as shown in Fig.~\ref{fig:marker-coordinate}. In this case, we make the detected marker the reference frame's new origin. All 3D interactions are performed locally as to this new reference frame and synchronized across different clients. By default, we render the volume on the marker when there is no 3D interaction (Fig.~\ref{fig:marker-coordinate}).

%--------------------------------------------------------------------
\section{User Interactions}
This section presents three types of user interaction methods under different tasks in the AR application. Fig.~\ref{fig:interaction-tasks} categorizes the interaction tasks and assigns the most intuitive interaction method to them.

\subsection{Virtual 2D Plane Touch}

We first created an xrHand system to achieve direct touch on a virtual 2D plane instead of pinch gesture as the "select" operation. The xrHand system actively tracks the right/left-hand index tip. We draw a small sphere around the tip and check the distance between the sphere to the target plane.

When drawing on a modern digital tablet, an intuitive stroke changes the brush size adaptively according to the pressure on the touch screen. The function that transforms the pressure into brush size is called pressure sensitivity. However, as the mid-air interaction doesn't have physical planes to provide actual pressure, we need to construct a virtual pressure mapping to gain pressure sensitivity. Our method is inspired by VirtualForce~\cite{ziyang2022virtualforce}, which proposed calculating the sketching force based on the difference between the user's physical hand position and their hand avatar in the Virtual Reality (VR) environment.

\begin{figure}
\centering
\begin{subfigure}[]{0.5\columnwidth}
\includegraphics[width=\columnwidth]{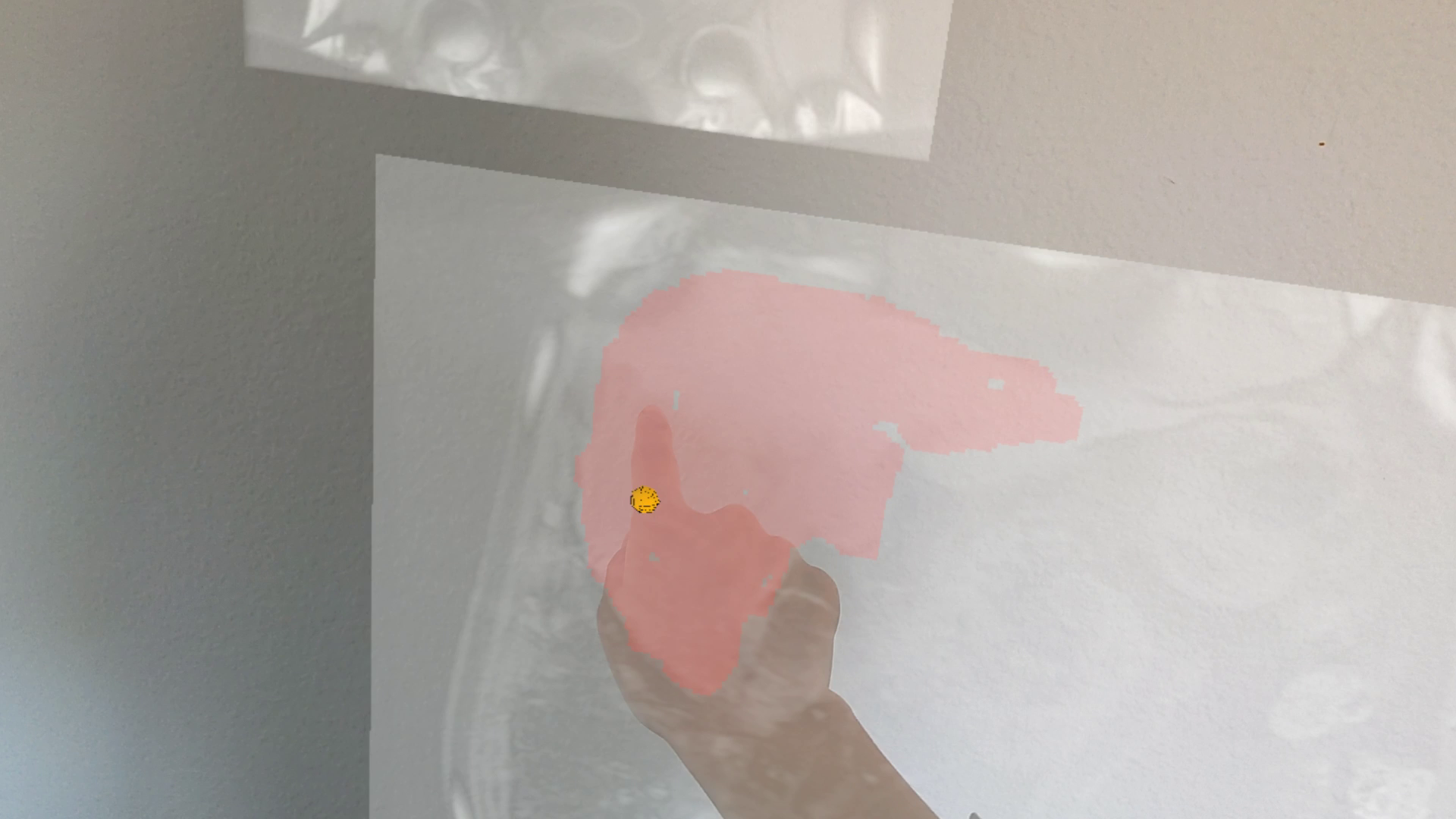}
\caption{}
\label{fig:virtual_force_screenshot-1}
\end{subfigure}%
\hfill
\begin{subfigure}[]{0.5\columnwidth}
\includegraphics[width=\columnwidth]{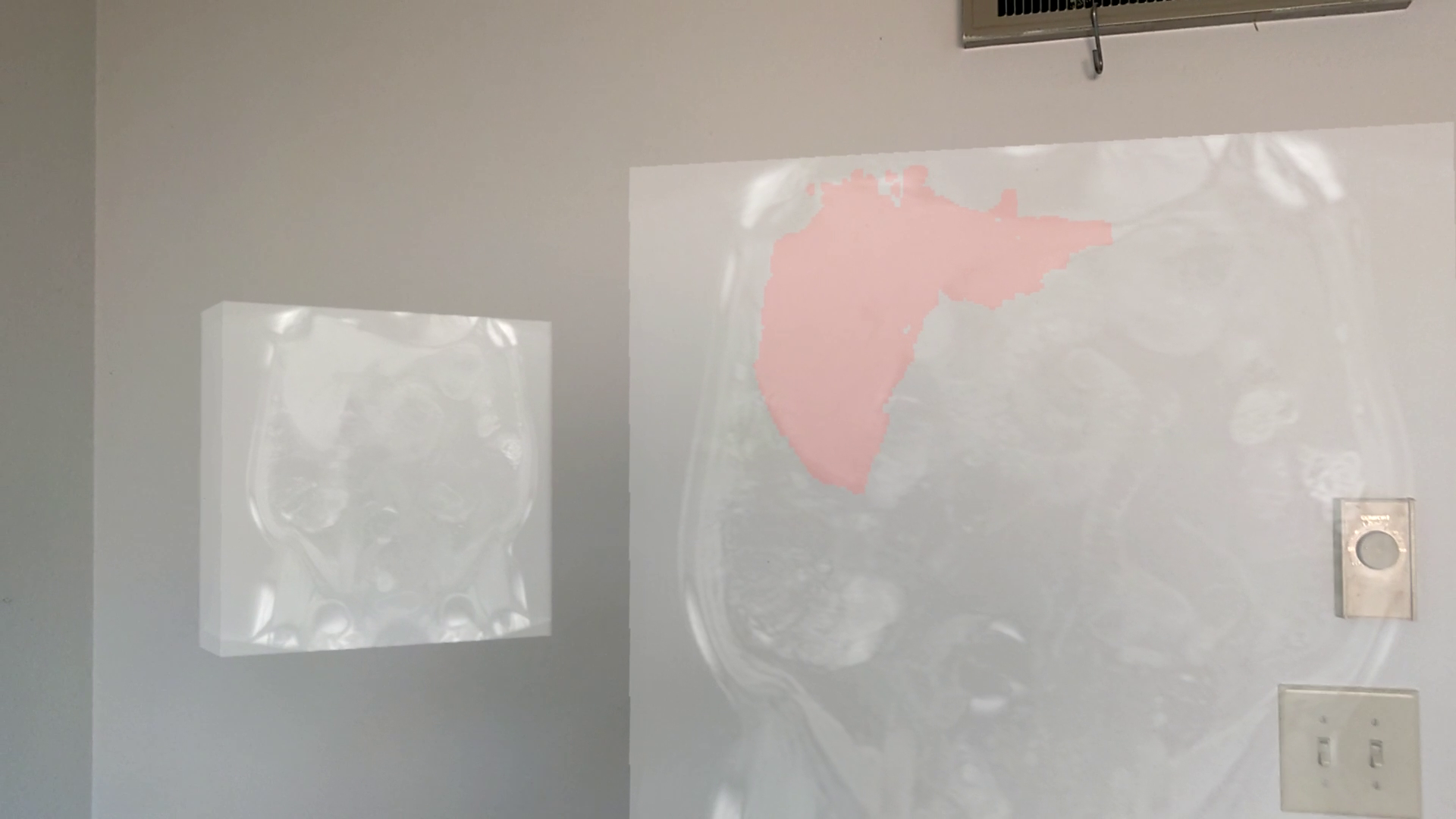}
\caption{}
\label{fig:virtual_force_screenshot-2}
\end{subfigure}%

\caption[]{Virtual-force sketch for 2D data annotation: (a) screenshot of 2D sketching; (b) side-by-side illustration of the volume and annotation data.}
~\label{fig:virtual_force_screenshot}
\end{figure}

% Fig.~\ref{fig:depth-force-scheme} illustrates our virtual-force sketching method: a
We transformed this concept into AR. As the index finger plunges into the virtual canvas, the hand-unaligned distance ($d$) determines the pen pressure on the canvas. The deeper the finger goes, the more pressure will be on the canvas (Equ.\ref{equ:depth-force}). Then with a linear pressure sensitivity function, we map the virtual pressure to the sketch stroke size for planar data annotation (Fig.~\ref{fig:virtual_force_screenshot}).

\begin{equation}
% \begin{split}
    d = planeNormal \cdot (P_r - planePoint),
    P_v = P_r - d * planeNormal 
% \end{split}
\label{equ:depth-force}
\end{equation}

\begin{figure}[ht!]

\begin{subfigure}[]{0.25\columnwidth}
\includegraphics[width=\columnwidth]{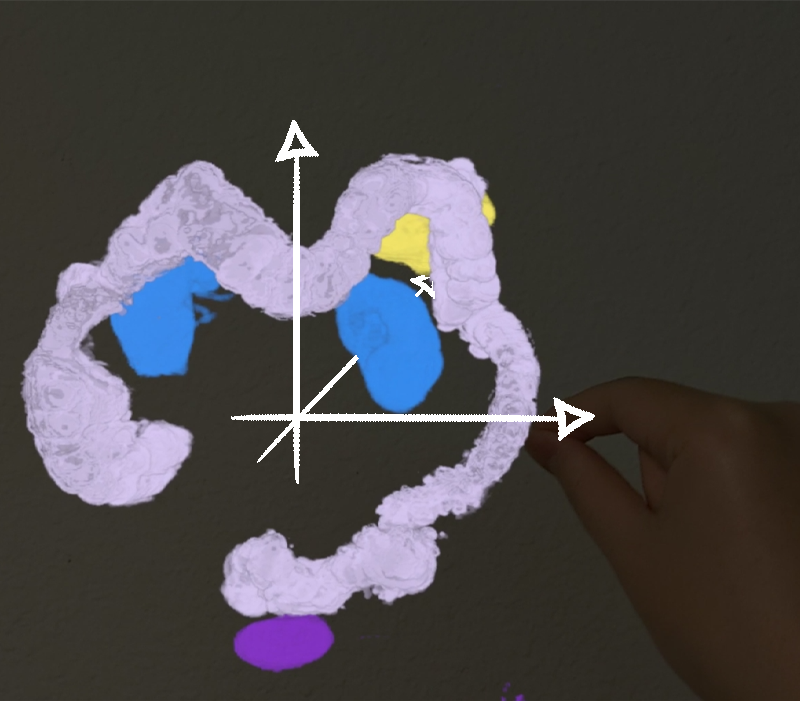}
\caption{}
\label{fig:3D_interaction-1}
\end{subfigure}%
%\hfill\vrule\hfill
\hfill
\begin{subfigure}[]{0.25\columnwidth}
\includegraphics[width=\columnwidth]{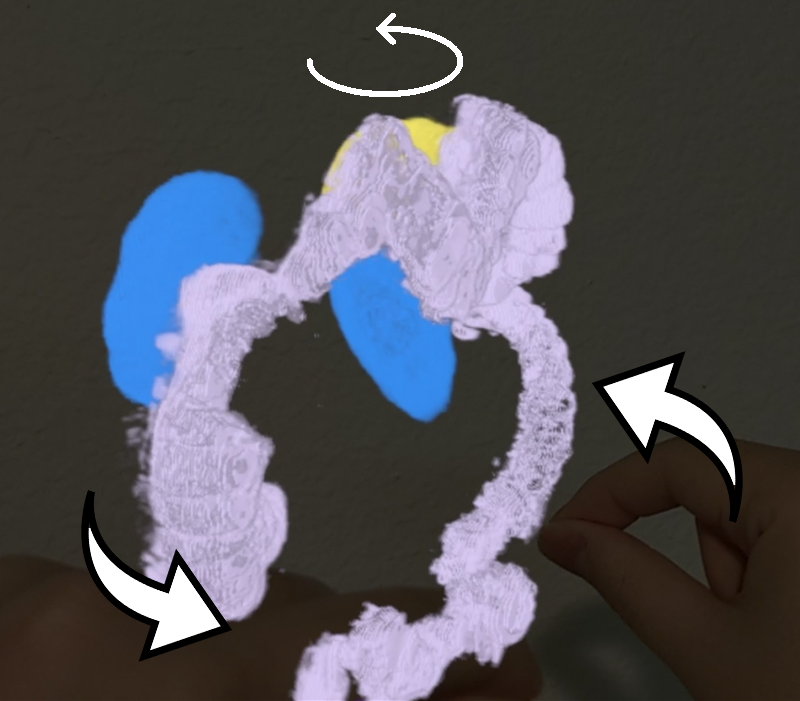}
\caption{}
\label{fig:3D_interaction-2}
\end{subfigure}%
\hfill
\begin{subfigure}[]{0.25\columnwidth}
\includegraphics[width=\columnwidth]{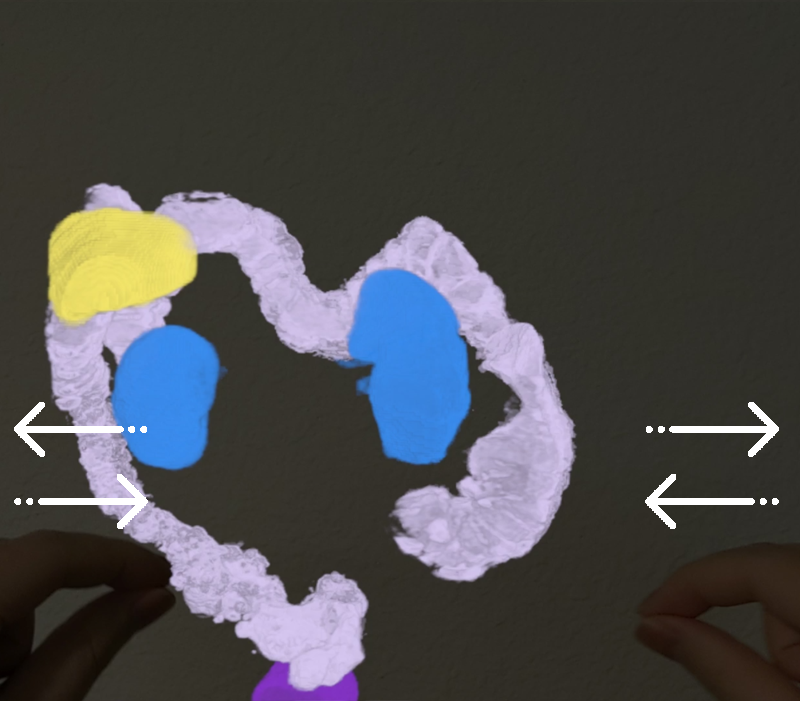}
\caption{}
\label{fig:3D_interaction-3}
\end{subfigure}%
\hfill
\begin{subfigure}[]{0.25\columnwidth}
\includegraphics[width=\columnwidth]{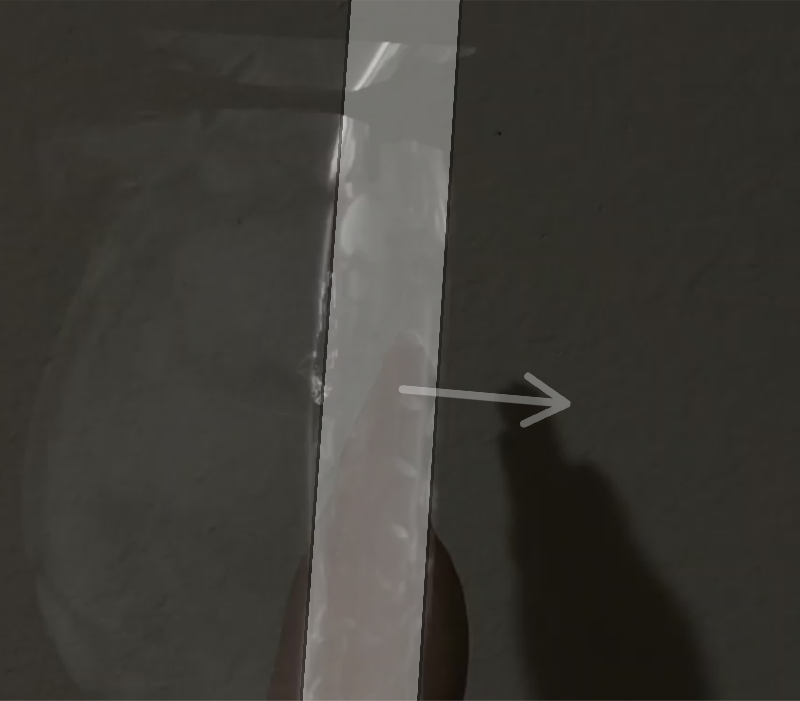}
\caption{}
\label{fig:3D_interaction-4}
\end{subfigure}%

\caption[]{Four different free-hand 3D interactions: (a) translation; (b) rotation; (c) scaling; (d) cutting plane}
~\label{fig:3D_interaction}
\end{figure}
\subsection{3D User Interaction}
We provide single-hand and two-hand gesture control for basic volume interactions and a cutting plane with 6 degrees of freedom. The user can manipulate the translation of the volume by firstly grabbing and then moving the single hand around as shown in Fig.~\ref{fig:3D_interaction-1}. With two hand gestures, the user can firstly hold the volume with both hands and then control the orientation and scale of the volume. Fig.~\ref{fig:3D_interaction-2} shows that grabbing and rotating two hands relatively will apply the same amount of rotation onto the volume. As shown in Fig.~\ref{fig:3D_interaction-3}, moving two hands away from each other enlarges the volume. Conversely, moving two hands closer will shrink the volume.
In the Fig.~\ref{fig:3D_interaction-4} we use the palm poses to control the cutting plane directions and depth.

%-------------------------------------------------------------------------
\section{Experiments}
We measured the frame rate while moving the volume on five scripted paths: rotating the volume around its y-axis and its x-axis, respectively, while being either $2$ m or $1$ m away from the user. Each of the rotations took 10 seconds. Furthermore, we moved the volume along the z-axis, starting again at $2$ m to $0$ m distance. This movement path required 10 seconds to complete. All frame rates were measured by averaging over 500 ms periods.

\begin{figure}
  \centering
  \includegraphics[width=\columnwidth]{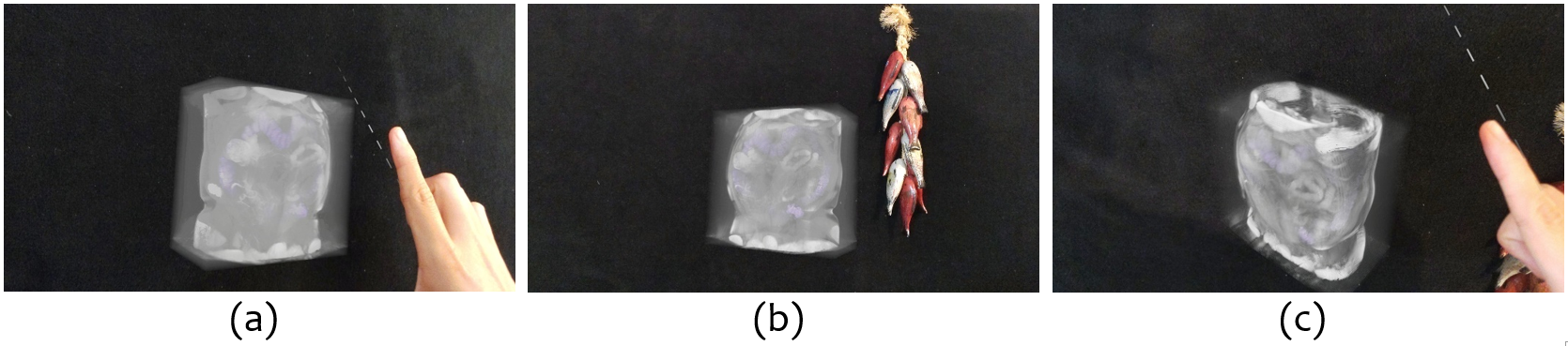} \caption{Screenshots of a volumetric data rendered with: (a) texture-based method, (b) view-aligned slicing method, and (c) raycasting method.}~\label{fig:three-method-rendering}
\end{figure}

The experiment volume contained $144$ scans, each of $512 \times 512$ pixels. The measurements are on three rendering methods with parameters as shown in Fig.~\ref{fig:three-method-rendering}.

\begin{figure}[h!]
\subfloat[A full rotation around the x-axis and y-axis setting at 1m and 2m distance]{%
\includegraphics[height=0.22\textheight]{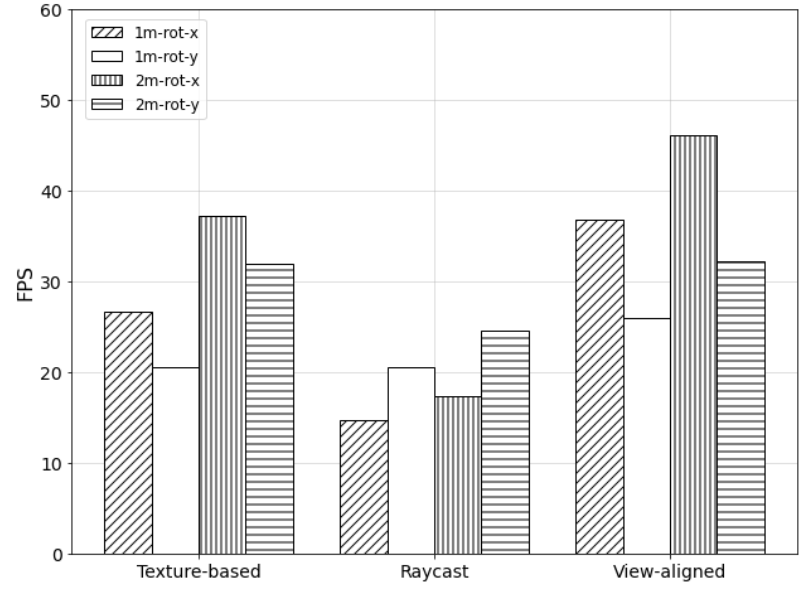}}
\hspace*{\fill}
\subfloat[A uniform rectilinear motion from 2m to 0m.]{%
\includegraphics[height=0.22\textheight]{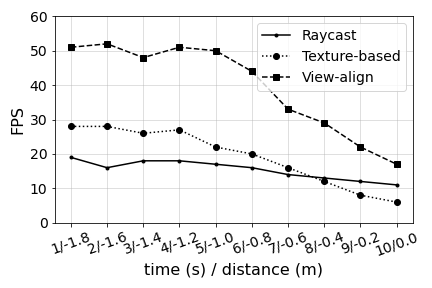}}
\caption{Experiment results: Average frame rates during (a)rotation and (b) translation.}
~\label{fig:experiments}
\end{figure}

In Fig.~\ref{fig:experiments}(a), we show the frame rates for the rotation around the x-axis and y-axis at $1$ m and $2$ m. View-aligned rendering methods reach the highest frame rate to 46 fps rotating around the x-axis at a 2 m distance. This method performs the best overall for reducing the number of slices that need blending to get a decent view yet fewer artifacts—followed by the texture-based rendering method, which manages to exceed 30 fps rotating at a 2m distance. The raycasting method is overall costly, although it accumulates its values adaptively. It performs better for a comparably static view scenario.

Fig.~\ref{fig:experiments}(b) shows the frame rates for uniform rectilinear motion from 2m to 0m away. The plot shows the frame rate change across the 2m movement in 10 seconds. For the View-aligned and Texture-based methods, from the beginning 2m to 1m, the frame rate drops slowly while it drops faster from 1m to 0m. As the volume moves closer, it fills the entire two views, making the alpha-blending much slower for reading from and writing to back buffer. The change of the Raycasting method is minor because instead of alpha-blending, it works within a compute shader to sample along a ray within limited steps.

%---------------------------------------------------------------------

\section{Conclusions}

In this paper, we presented a prototypical Mixed Reality DICOM viewer based on a DirectX-based native engine. The functionalities of this prototype match a previously developed DICOM Viewer-Android system, while optimizing for two-views head-mount displays. The matched functionalities make it possible and intuitive to remote control the view inside a HoloLens with a connected smartphone. It also enables new possibilities, such as supporting collaborative work across platforms. Our testing results shows that it is difficult to achieve 60 fps rendering on two views simultaneously while maintaining a high-quality rendering for a volume. However, by switching among different rendering methods while adjusting the parameters, one can find the best balance between frame rate and quality.
%
% ---- Bibliography ----
%
% BibTeX users should specify bibliography style 'splncs04'.
% References will then be sorted and formatted in the correct style.
%
\bibliographystyle{splncs04}
% \bibliography{mybibliography}
%
% \bibliographystyle{sn-mathphys} 
\bibliography{ref}
\end{document}